\documentstyle[epsfig, twocolumn]{iso98}

\newcommand{\thepapername}{}

\def\hii{H~{\sc ii}}

\def\neii{[Ne~{\sc ii}]}
\def\neiii{[Ne~{\sc iii}]}

\def\siii{[S~{\sc iii}]}
\def\siv{[S~{\sc iv}]}
\def\arii{[Ar~{\sc ii}]}
\def\ariii{[Ar~{\sc iii}]}
\def\micron{$\mu$m}

\def\msun{\ifmmode M_{\odot} \else M$_{\odot}$\fi}

\def\teff{$T_{\rm eff}$}

\def\costar{{\em CoStar}}

\begin{document}

\setlength{\parindent}{0pt}
\setlength{\parskip}{ 10pt plus 1pt minus 1pt}
\setlength{\hoffset}{-1.5truecm}
\setlength{\textwidth}{ 17.1truecm }
\setlength{\columnsep}{1truecm }
\setlength{\columnseprule}{0pt}
\setlength{\headheight}{12pt}
\setlength{\headsep}{20pt}
\pagestyle{veniceheadings}

\title{\bf THE IMPACT OF NEW IONIZING FLUXES ON ISO OBSERVATIONS OF
  HII REGIONS AND STARBURSTS
\thanks{To appear in ``The Universe as
  seen by ISO'', Ed.\ P. Cox, M.F. Kessler, ESA Special Publications
series (SP-427)}
}

\author{{\bf Daniel Schaerer$^1$, Gra\.{z}yna Stasi\'nska$^2$} \vspace{2mm} \\
$^1$Observatoire Midi-Pyr\'en\'ees, Laboratoire d'Astrophysique, Toulouse,
France;
schaerer@obs-mip.fr \\
$^2$DAEC, Observatoire de Meudon, Meudon, France; grazyna.stasinska@obspm.fr }

\maketitle
\vspace*{-2cm}

\begin{abstract}
Extensive grids of photoionization models have been calculated for single star
\hii\ regions and evolving starbursts. We illustrate the predictions for
IR fine structure lines which are used to analyse the stellar content, and
derive properties such as the age and IMF.
The impact of recent ionizing fluxes on the IR lines are shown.
First comparisons of our starburst models with IR-diagnostics and the
ISO observations of Genzel et al.\ (1998) are also presented.
  \vspace {5pt} \\

  Key~words: HII regions, galaxies: starburst, ISO.

\end{abstract}

\section{INTRODUCTION}
The analysis of fine structure lines is crucial to interpret IR observations
of ultra-compact \hii\ regions, giant \hii\ regions and starburst galaxies
in terms of their stellar content. Through photoionization modeling the
nature of the ionizing sources of \hii\ regions and properties of starburst
regions such as the age and the IMF (slope, upper mass-cutoff) can in principle
be determined. Considerable progress has been made in the recent year in this
respect through the use of new stellar fluxes, spectral energy distributions
appropriate for evoling stellar populations at different metallicities and
including detailed observational constraints (e.g.\ Garcia-Vargas et al.\ 1995,
1997, Stasi\'nska \& Leitherer 1996, Stasi\'nska \& Schaerer 1997,
Gonz\'alez-Delgado et al. 1998).
None of these studies have, however, focussed on IR observation which are
now becoming widely available thanks to ISO.

In this contribution we present exploratory results regarding predictions of
IR lines from two sets of recent photoionization models:
{\em 1)} Single star \hii\ regions using the latest ionizing fluxes from
non-LTE, line blanketed atmospheres including stellar winds (\costar\ models,
Schaerer \& de Koter 1997) and,
{\em 2)} Photoionization models for evolving starburst populations based on the
Leitherer \& Heckman (1995) synthesis models.
Detailed results from new photoionization models for starbursts including the
\costar\ models and updated stellar tracks will be presented elsewhere.

\section{SINGLE STAR HII REGION MODELS}
In Figure 1 we illustrate the behaviour of the line ratios of some
important fine structure lines observable in the SWS range and with
ISOCAM/CVF.
This figure, based on the model calculations of Stasi\'nska \& Schaerer (1997,
hereafter SS97), shows the
dependence of the line ratios on the stellar temperature and --
importantly -- also on the ionization parameter.
The comparison between the top and bottom row shows the
differences obtained using either the recent \costar\ atmosphere
models or Kurucz plane parallel LTE models.
Due to the harder spectrum predicted by the \costar\ models
above $\sim$ 35-40 eV (Schaerer \& de Koter) a {\em higher excitation} is
generally
predicted for a given \teff\ than with Kurucz models (see SS97).

Evidence supporting such an increase of the hardness of the ionizing
flux comes from the KAO observations of Rubin and coworkers (cf.\
Simpson et al. 1995). Indeed, the \costar\ models including stellar
winds, non-LTE effects and line blanketing are thereby able to
provide a satisfactory description of the Ne$^{++}$/O$^{++}$ ionic
fractions (see SS97, also Sellmaier et al.\ 1996).

\begin{figure*}[!ht]
\centerline{\epsfig{figure=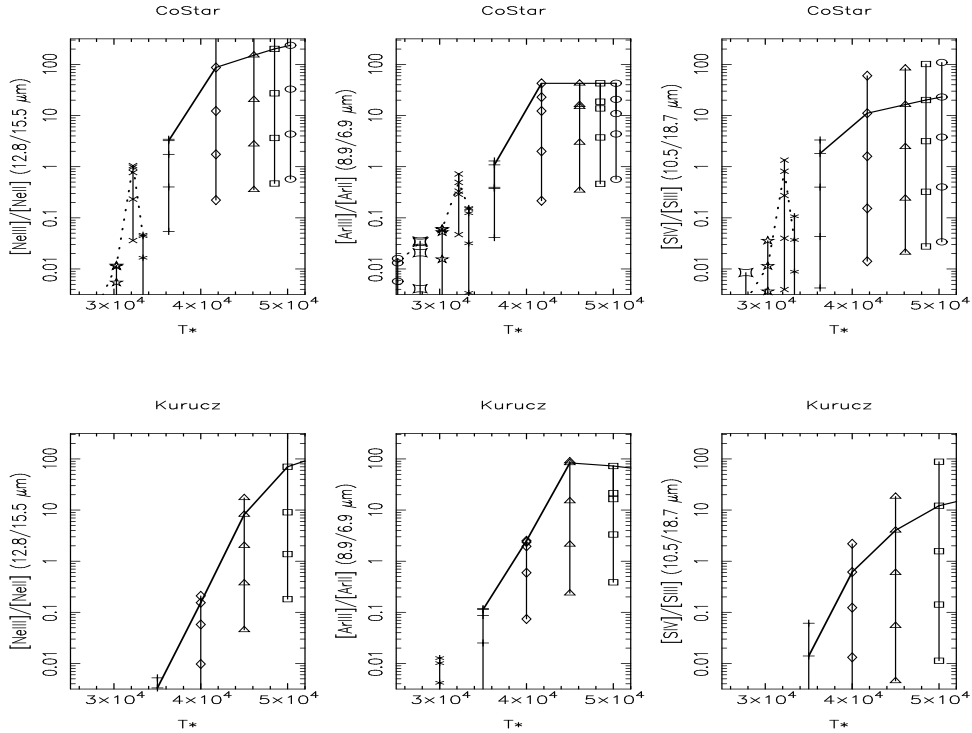,height=10.5cm,width=13cm}}
 \caption{\em Top row: Photoionization models using \costar\ fluxes
(dwarf models for
        \teff $>$ 35000 K, supergiant models otherwise.
        Bottom: same with Kurucz models.
        Models at solar metallicity with varying ionization parameter
        $U$ over 4 dex (typically $\log U \sim$ -5 to -1).
        Solid lines connect models with $U$  typical for giant \hii\
        regions or for bright single star \hii\ regions (``reference
model'' of SS97).}
\end{figure*}

Encouraged by these successes atmosphere models including the most
relevant physical processes (stellar winds, non-LTE, blanketing)
continue to be improved. See e.g.\ Pauldrach (1997) and Schaerer
(1997) for recent progress reports.
In addition more stringent tests using ISO observations of nebulae
with well known ionizing sources would be extremely valuable.
Detailed photoionization modeling including constraints on the nebular
geometry and likely also their dust content will be necessary to
achieve such a goal.

In Fig.\ 2 we plot the predictions for single star \hii\ regions for some
quantities of interest for the interpretation of integrated populations
discussed later.
The middle and right panels show the ratio of luminosity in the Lyman continuum
(assuming an average photon energy of 16 eV) $L_{\rm Lyc}$ to the \neii\ and
\siii\ line luminosity respectively.
It is useful to understand the behaviour of the \neii/\siii\ 12.8/33.5
ratio (left) since either of these lines is sometimes used as a proxy in
case the
other line is unobservable (cf.\ Genzel et al.\ 1998).
Indeed for a given \teff\ one obtains a relatively small variation ($\sim$
1 dex)
of \neii/\siii\ with the ionization parameter.
In  all our  ``reference models'' (constructed assuming a gas density of 10
cm$^{-3}$)
we predict \neii/\siii\ $<$ 1.
For stars with \teff\ $>$ 36 kK the differences between the \costar\ and Kurucz
fluxes become non-negligible for \neii/\siii.
Some implications of these finding are briefly discussed below.

\protect\vspace*{-0.4cm}
\section{PHOTOIONIZATION MODELS FOR EVOLVING STARBURSTS}
IR fine structure lines provide a new tool to study the hot star
population of obscured starbursts. Through comparisons with photoionization
modeling this allows in particular to constrain the star formation history
and the upper end of the initial mass function (e.g.\ Lutz et al.\ 1998).
Empirical diagnostic diagrams based on IR lines also allow the distinction
between
starburst and AGN powered sources (Genzel et al.\ 1998).
Theoretically these diagnostics remain, however, to be understood and
quantified.

To address such questions photoionization models representing evolving
starbursts at different metallicities are nowadays required.
In the following we will briefly illustrate some predictions from such
models for
fine structure lines in the ISO LWS range. The predictions shown here are
drawn from
the models of Stasi\'nska \& Leitherer (1996, hereafter SL96,
available by anonymous ftp from ftp.obspm.fr/pub/obs/grazyna/cd-crete).
Detailed results from new photoionization models for young starbursts based on
the most recent evolutionary synthesis models of Schaerer \& Vacca (1998)
including
the latest Geneva stellar tracks and \costar\ model atmospheres (cf.\
above) will
be presented elsewhere.

In Figure 3 we show the computed temporal evolution of the line ratios of
\neiii/\neii\ 12.8/15.5
\micron, \ariii/\arii\ 8.9/6.9 \micron, and \siv/\siii\ 10.5/18.7 \micron\
(all top row)
for an instantaneous burst at 1/4 solar metallicity with a
Salpeter IMF and $M_{\rm up} =100$ \msun\ surrounded by a gas cloud
of density n=10 cm$^{-3}$. Different symbols represent
models with different (initial) ionization parameters: typically one has
$\log U \sim -5$ (circles) to $\log U \sim -1$ (crosses).
Note that in photoionization models of evolving starbursts, the dependence of
$U$ with time is a function of the adopted geometry, as stressed by
Stasi\'nska (1998). In the models of  SL96, $U$ decreases slightly
with time.

As expected the excitation decreases with the age of the burst.
It must, however, be reminded that the considered lines also strongly depend
on the ionization parameter (see Fig.\ 3).
Indeed the line ratios shown here vary typically by three orders of magnitudes
for variations of $U$ of 4 orders of magnitudes.
For comparison a change of the upper mass-cutoff from 50 to 100 \msun\
increases the same line ratios by less than one order of magnitude
(e.g.\ Lutz et al.\ 1998).
In general $U$ is poorly constrained.
Although optical spectra of \hii\ galaxies can be reproduced with models
covering only a fairly small range of $U$ ($\sim$ 1 dex, SL96) this is
not necessarily the case for ISO observations of starburst galaxies.
This question needs a careful analysis before meaningful results
can be derived on the IMF, age etc.

Genzel et al.\ find an empi\-ri\-cal va\-lue of \\
\neii/\siii\ $=$ 1.7 $\pm$ 1.3 (or 1.2 $\pm$ 0.6 excluding the two largest
outliers)
from their sample of 10 starbursts. The SL96 models constructed with a
density of n=10 cm$^{-3}$
typically predict \neii\ $<$ \siii\
 (this is true also for solar metallicity models, not shown here). However,
the SL96 models
with  n$=10^4$ cm$^{-3}$ reach \neii/\siii\ of 1.5, due to collisional
deexcitation of \siii.
The electron densities deduced from \siii\ 18.7/33.5 
for a subset of starburst galaxies from the Genzel et al.\ sample
are typically $< 10^{2-2.5}$,
so it seems that collisional deexcitation is not sufficient to explain the large
\neii/\siii\ observed.
Among possible factors to reduce the discrepancy are an Ne overabundance
(which however  needs to be understood in more detail), or a softer ionizing radiation field.
We conclude that the empirical value \neii/\siii $\sim$ 1.7 for starbursts 
used in the diagnostic diagrams of Genzel et al.\ (1998) is theoretically not 
well understood yet.

For comparisons of the Lyman continuum luminosity $L_{\rm Lyc}$ to the
bolometric
luminosity of starbursts, ULIRGS, etc.\ it is interesting to be able to derive
$L_{\rm Lyc}$ from the IR fine structure lines. This requires, however, some
empirical calibration or the use of models.
``Empirical'' values of $L_{\rm Lyc}/L($\neii$) = 64 \pm 37$ and
$L_{\rm Lyc}/L($\siii$) = 105 \pm 61$ are obtained from the starburst sample of
Genzel et al.\ excluding NGC 5253, which shows much larger values.
It must be reminded that $L_{\rm Lyc}$ is derived from the Brackett
recombination
lines assuming case B and an average photon energy of 16 eV, and neglecting the
absorption of Lyc photons by dust.

Our model predictions show in Fig.\ 3 (mid/right bottom panels) for the
instantaneous
burst model show quite a wide range for the $L_{\rm Lyc}/$line ratios,
depending
on age and ionization parameter.
Although these idealised models (instantaneous burst) are not necessarily
representative for the average starburst population observed through the ISO
aperture it is instructive to compare our predictions with the ``observed''
values of Genzel et al.
In any case the predicted values of  $L_{\rm Lyc}/L($\neii$)$ are
systematically
larger than the empirical starburst value even for the oldest age considered
here.
Using the \costar\ model atmospheres will even further increase the
predicted value
given their harder energy distribution with respect to the Kurucz ones.
%
The $L_{\rm Lyc}/L($\siii$)$ ratio shows a better agreement with the average 
value from Genzel et al.\ This is consistent with the discrepancy between the 
model predictions for \neii/\siii\ and the observations discussed earlier.
Most probably, the Lyc luminosities derived from the observations are
underestimated because they do not take into account dust absorption of Lyc photons.
More detailed analysis are required to obtain consistent a consistent picture
of the fine structure lines in such objects.

\vspace{-0.5cm}
\section{CONCLUSIONS}
We have illustrated the impact of the recent \costar\ ionizing fluxes on
IR fine structure lines for single star \hii\ regions and shown predictions
of IR lines for evolving starbursts resulting from calculations of extensive
grids of photoionization models (see SL96, SS97).

As shown earlier, ionizing fluxes from non-LTE, line blanketed atmospheres
including stellar winds predict a harder spectrum (Schaerer \& de Koter 1997)
leading to higher excitation in the nebula. These predictions are supported
by comparisons with optical and pre-ISO (KAO) IR observations of \hii\ regions
(SS97 and references therein).
Detailed models of ultra-compact \hii\ regions
or \hii\ regions with well-known stellar content based on ISO observations
and other observational constraints will be extremely useful to better
constrain the atmosphere models and probe the importance of other effects
(e.g.\ dust) usually neglected in photoionization models. Such studies
also provide important tests for our understanding of more complex objects
like starbursts.

For starbursts we have illustrated the behaviour of line ratios which can
in principle be used to constrain burst properties such as the age, IMF slope
and upper mass-cutoff.  We have also studied the behaviour of the
\neii/\siii\ ratio
and diagnostics of $L_{\rm Lyc}$ from fine structure lines extensively used by
Genzel et al.\ (1998).
We have shown that a thorough analysis of the  properties of the
{\em nebular} gas (ionization parameter, geometry, dust content, total
amount of gas)
is needed
for reliable determinations of the {\em stellar} content of the nebulae
observed by ISO.

\vspace{-0.2cm}
{\small
{\em ACKNOWLEDGMENTS} 
DS thanks the LOC for financial support.
}


\begin{figure*}[!hb]
\centerline{\epsfig{figure=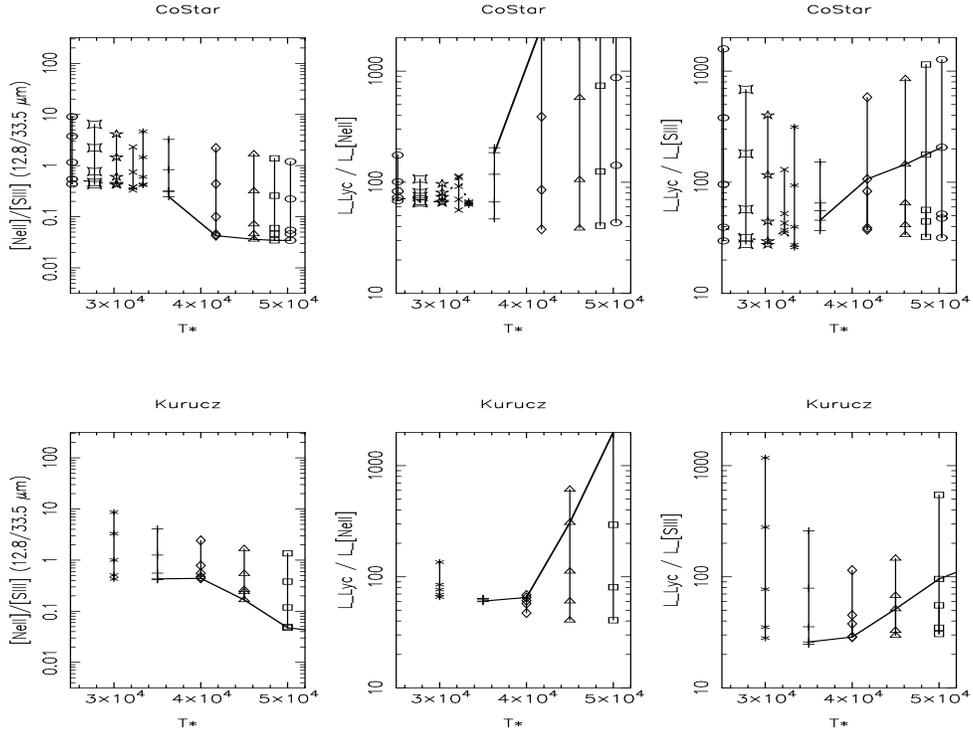,height=10.5cm,width=13cm}}
\caption{\em  Same conventions as Fig.\ 1.}
\end{figure*}

\vspace*{-2cm}
\begin{figure*}[htb]
\centerline{\psfig{figure=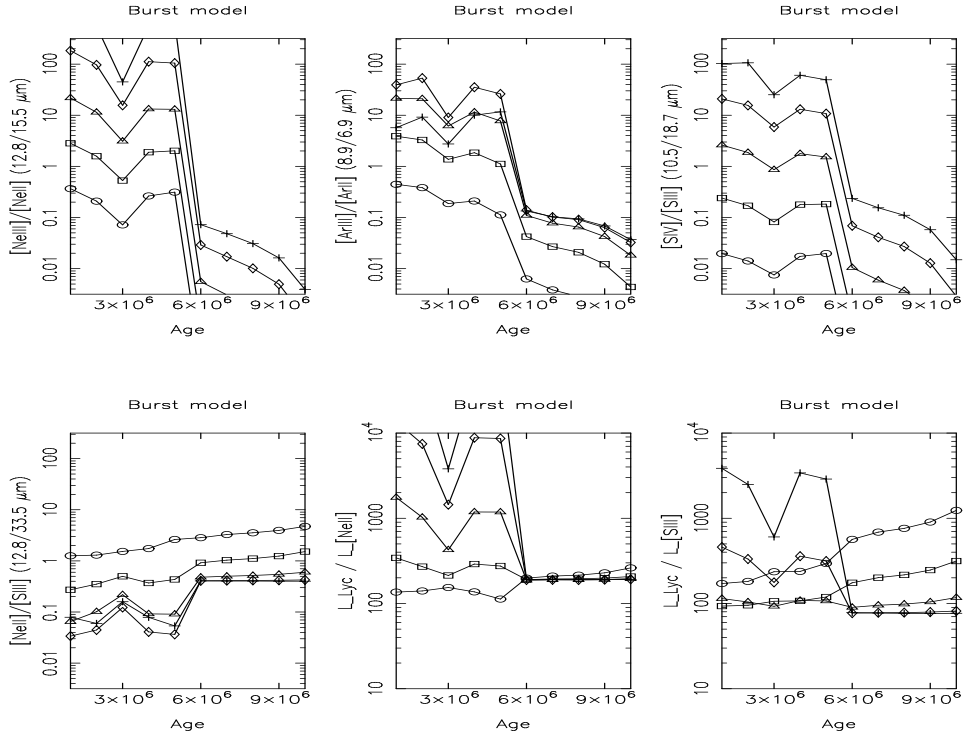,height=10.5cm,width=13cm}}
 \caption{\em Photoionization models for evolving starburst
   (instantaneous burst, Salpeter IMF, $M_{\rm up}=$100 \msun,
   $Z=1/4 Z_\odot$) and varying ionization parameters
   ($\log U \sim$ -5 to -1, circles to crosses).}
\end{figure*}

\end{document}